\documentclass[copyright,creativecommons]{eptcs}
\providecommand{\event}{CREST 2018} 
\usepackage{breakurl}             
\usepackage{underscore}           

\usepackage{amssymb,amsmath}
\usepackage[mathscr]{eucal}
\usepackage{verbatim}

\usepackage{paralist}
\usepackage{mathtools}
\usepackage{amsfonts}
\usepackage{color}
\usepackage{graphicx}
\usepackage{tikz}
\usepackage{xspace} 

\usepackage[usenames,dvipsnames]{pstricks}
\usepackage{epsfig}
\usepackage{pst-grad} 
\usepackage{pst-plot} 
\usepackage{pstricks-add}

\DeclareMathOperator{\myinp}{\mathsf{inp}}

\DeclareMathOperator{\myout}{\mathsf{out}}
\DeclareMathOperator{\tr}{\mathsf{tr}}
\newcommand{\cset}{\mathcal{D}}

\title{Causality Analysis for Concurrent Reactive Systems\\(Extended Abstract)}
\author{Rayna Dimitrova
\institute{University of Leicester\\Leicester, UK}
\email{rd307@leicester.ac.uk}
\and
Rupak Majumdar
\institute{MPI-SWS\\
Keiserslautern and Saarbr\"ucken, Germany}
\email{rupak@mpi-sws.org}
\and
Vinayak S.\ Prabhu
\institute{Colorado State University\\
Fort Collins, Colorado, United States }
\email{vinayak@mpi-sws.org}
}
\def\titlerunning{Causality Analysis for Concurrent Reactive Systems}
\def\authorrunning{R. Dimitrova, R. Majumdar \& V. S. Prabhu}

\begin{document}
\maketitle

\begin{abstract}
We present a comprehensive language theoretic causality analysis 
framework for explaining safety property violations 
 in the setting of concurrent reactive systems. 
Our framework allows us to uniformly express a number of causality notions studied in the areas of artificial intelligence and formal methods, as well as define new ones that are of potential interest in these areas. Furthermore, our formalization provides means for reasoning about the relationships between individual notions which have mostly been considered independently in prior work; and allows us to judge the appropriateness of the different definitions 
 for various  applications in  system design.
In particular, we consider causality analysis notions for debugging, error resilience,
and liability resolution in concurrent reactive systems.
Finally, we  present automata-based algorithms for computing various causal sets based on our
language-theoretic encoding, and derive the algorithmic complexities.
\end{abstract}

Causality analysis, which investigates questions of the form ``Does event $e_1$ cause event $e_2$?'' 
plays an important role in many areas of 
science, medicine and law.
In formal methods, causality analysis has been used 
to determine the \emph{coverage of specifications}~\cite{ChocklerHK08} (that is, which parts of the system under scrutiny are relevant for the satisfaction of a specification),
to \emph{explain counterexamples}~\cite{BeerBCOT12} (identify points in a counterexample trace that are relevant for the failure of a temporal specification),
to \emph{construct fault trees}~\cite{Leitner-FischerL13},
and to \emph{automatically refine system abstractions}~\cite{ChocklerGY08}.
In artificial intelligence, causality-based explanation finding has applications in natural language processing, automated medical diagnosis, vision processing, and planning.
Resolving liabilities in a legal setting often relies on establishing the causal relations between potential causes and the occurred damage~\cite{EuropeanTortLaw}.

Causality definitions based on \emph{counterfactuals}, which are alternative scenarios where the suspected cause $e_1$ of $e_2$ did not happen, date back to~\cite{Hume1748} and 
have been extensively studied in philosophy~\cite{Lewis2004}. 
In computer science, the most prominent and widely used definition of causality is that 
of~\cite{HalpernP05}, in which the authors write
``... while it is hard to argue that our definition (or any other definition, for that matter) is 
the right definition, we show that it deals with the difficulties that have plagued other approaches in the past ...''.
Halpern and Pearl's 
approach is based on \emph{structural equations}, which describe  causal dependencies between Boolean variables.
We extend the Boolean study of causality to the \emph{temporal} setting; specifically, we formalize notions of causality in 
\emph{concurrent reactive systems} whose behaviors evolve over time.
A concurrent reactive system is a
 composition of interacting components; the system behavior is
 determined by the \emph{repeated} interaction between the components over time.
Moreover, we consider the setting where component implementations are not available for analysis and the designer has only access to specifications of their expected behavior. 
Thus,
when analyzing an \emph{error trace} (an execution of the system that violates a desired system-level property), 
the only available information about the system consists of the components' specifications and the observed trace.

In our framework, a concurrent reactive system $C_1 \parallel C_2 \parallel \ldots \parallel C_n$ is
a composition of components $C_1, \ldots, C_n$. 
Each component $C_i$ is  specified as a tuple  $(X_i, \myinp(X_i), \myout(X_i), \Sigma_i, \varphi_i)$, where
\begin{compactitem}
\item 
$X_i = \myinp(X_i) \uplus \myout(X_i)$
is the set of variables of the component, 
consisting of the input variables $\myinp(X_i)$ and 
the output variables $\myout(X_i)$ (the sets of input and output
variables being disjoint);
\item 
$\Sigma_i$ is the alphabet, 
consisting of all possible valuations of the variables $X_i$;
\item $\varphi_i$ is a non-empty prefix-closed language over $\Sigma_i$, specifying the set of correct behaviours of $C_i$.
\end{compactitem}
The composite system $C_1 \parallel C_2 \parallel \ldots \parallel C_n$ has an associated prefix-closed
specification $\theta$ such that $\theta$ contains $\varphi_1 \parallel\ldots \parallel \varphi_n$.
Thus, the global requirement is more relaxed than the
promised behaviors of the individual components.
In other words, the system $C_1 \parallel C_2 \parallel \ldots \parallel C_n$ promises to
implement or refine the global requirement
$\theta$.

Consider a trace $\tr$ of the system  $C_1 \parallel C_2 \parallel \ldots \parallel C_n$
in which the system requirement $\theta$ is violated.
Let $C'_1, \ldots, C'_k$ be the components which violate their local specifications
$\varphi_1', \ldots, \varphi_k'$.
The \emph{causality analysis problem} is to determine which component set
$\{D_1, \ldots, D_m\} \subseteq \{C'_1, \ldots, C'_k\}$ is liable for the global system requirement
violation in $\tr$.
Our analysis reasons about  two classes  of scenarios
to determine if   $\cset= \{D_1, \ldots, D_m\}$  is a cause:
\begin{compactitem}
\item 
\emph{Fault Mitigation Capability} analysis asks
whether  the  \emph{correct behavior} of the components in the set $\cset$ is enough
to mitigate the faults of all components (\emph{including those of components not in $\cset$}), 
by ensuring that the required system property holds.
\item 
\emph{Fault Manifestation} analysis asks
whether the observed \emph{faulty behavior} of the components in the set $\cset$ is 
enough to 
manifest a global fault (\emph{i.e.}, a system behavior violating the global property),
even if the components not  in $\cset$ were to behave correctly.
\end{compactitem}
These two classifications parallel the classifications of~\cite{GosslerMR10,GosslerM13}
 of causes into
\emph{necessary causes} and \emph{sufficient causes}.
However, our analysis is not limited to specific definitions of counterfactual sets. In contrast, we provide a reasoning framework based on generic counterfactual sets, and introduce several natural instantiations.
 We demonstrate that the generality and  modularity of our definition of causality allow us to seamlessly extend causality analysis to the case of \emph{heterogeneous fault models}, where different components are examined under different fault scenarios.
Finally, we present an automata-based method for determining various causal sets
in the setting of  heterogeneous component-fault models,
 and derive its algorithmic complexity.

\bibliographystyle{eptcs}
\bibliography{ref}

\end{document}